\documentclass[aps,prl,twocolumn,superscriptaddress, fleqn, showpacs]{revtex4}

\usepackage{times,xspace}
\usepackage{amsbsy,amssymb,amsmath,bm}
\usepackage[pdftex]{graphicx,color}
\usepackage{longtable}

\def\bbbc{{\mathchoice {\setbox0=\hbox{$\displaystyle\rm C$}\hbox{\hbox
to0pt{\kern0.4\wd0\vrule height0.9\ht0\hss}\box0}}
{\setbox0=\hbox{$\textstyle\rm C$}\hbox{\hbox
to0pt{\kern0.4\wd0\vrule height0.9\ht0\hss}\box0}}
{\setbox0=\hbox{$\scriptstyle\rm C$}\hbox{\hbox
to0pt{\kern0.4\wd0\vrule height0.9\ht0\hss}\box0}}
{\setbox0=\hbox{$\scriptscriptstyle\rm C$}\hbox{\hbox
to0pt{\kern0.4\wd0\vrule height0.9\ht0\hss}\box0}}}}

\newcommand{\ybal}{{$\beta$-YbAlB$_4$}} 
\newcommand{\lbal}{{$\beta$-LuAlB$_4$}}

\begin{document}

%Title of paper
\title{The role of \textit{f}-electrons in the Fermi surface of the heavy fermion 
superconductor $\beta$-YbAlB$_4$} 

%\author{author list}
\author{E. C. T. O'Farrell}
\affiliation{Cavendish Laboratory, University of Cambridge, Madingley Road,
  Cambridge CB3 0HE, UK}
\author{D. A. Tompsett}
\affiliation{Cavendish Laboratory, University of Cambridge, Madingley Road,
  Cambridge CB3 0HE, UK}
\author{S. E. Sebastian}
\affiliation{Cavendish Laboratory, University of Cambridge, Madingley Road,
  Cambridge CB3 0HE, UK}
\author{N. Harrison}
\affiliation{NHMFL, MS-E536, Los Alamos National Laboratory, Los Alamos, New
  Mexico 87545, USA}
\author{C. Capan}
\affiliation{Department of Physics and Astronomy, University of California,
  Irvine, California 92697-4575}
\author{L. Balicas}
\affiliation{National High Magnetic Field Laboratory, Tallahassee, Florida
  32310, USA}
\author{K. Kuga}
\affiliation{Institute for Solid State Physics, University of Tokyo, Kashiwa,
  Japan 277-8581}
\author{T. Matsuo}
\affiliation{Institute for Solid State Physics, University of Tokyo, Kashiwa,
  Japan 277-8581}
\author{M. Tokunaga}
\affiliation{Institute for Solid State Physics, University of Tokyo, Kashiwa,
  Japan 277-8581}
\author{S. Nakatsuji}
\affiliation{Institute for Solid State Physics, University of Tokyo, Kashiwa,
  Japan 277-8581} 
\author{G. Cs\'{a}nyi}
\affiliation{Department of Engineering, Cambridge University, Trumpington
  Street, Cambridge CB2 1PZ, UK}
\author{Z. Fisk}
\affiliation{Department of Physics and Astronomy, University of California,
  Irvine, California 92697-4575}
  \author{M. L. Sutherland}
\affiliation{Cavendish Laboratory, University of Cambridge, Madingley Road,
  Cambridge CB3 0HE, UK}

\date{\today}
\begin{abstract}

%Brief summary of what was done:

We present a detailed quantum oscillation study of the Fermi surface of the 
recently discovered Yb-based heavy fermion superconductor $\beta$-YbAlB$_4$. 
We compare the data, obtained at fields from 10 to 45 Tesla, to band structure 
calculations performed using the local density approximation. Analysis of the data 
suggests that $f$-holes participate in the Fermi surface up to the highest magnetic fields studied. We comment on the significance of these findings for 
the unconventional superconducting properties of this material.

\end{abstract}

\pacs{71.18.+y, 71.20.Eh, 71.27.+a, 75.30.Mb, 75.10.Lp}

\maketitle

The recent demonstration of superconductivity in the heavy fermion compound \ybal{} \cite{Nakatsuji08} provided a breakthrough in the search for a ytterbium-based analogue of the numerous cerium heavy fermion
superconductors \cite{Steglich79,Petrovic01,Settai07}. Theoretical models of $f$ electron metals
often describe the low temperature dynamics in terms of the dense
Kondo lattice hamiltonian \cite{VonLohneysen07}. In this framework, it is expected that hole and electron like materials should realize similar ground states. While Yb in a $4f^{13}$ electronic configuration is the hole 
counterpart to $4f^1$ Ce, superconductivity in Yb-based materials \cite{Gegenwart02,Knebel06} has proved strangely elusive in contrast to those containing Ce, until the discovery of superconductivity in \ybal.

Single crystals of \ybal{} grow in a plate-like morphology, with a crystal structure corresponding to that
of ThMoB$_4$ \cite{Macaluso07}. In contrast
to the more common square Yb superlattice, the Yb ions in \ybal{} form a lower symmetry distorted hexagon, with Yb and Al atoms in the basal ab-plane sandwiched between nets containing an unusual arrangement of heptagonal and pentagonal B rings. Alloying with smaller interstitial atoms favours a smaller 4$f^{13}$ ionic configuration with the shortest Yb-Yb bond length lying along the c-axis. Experimental studies of the magnetic susceptibility \cite{Nakatsuji08} show a large Ising-like moment along the easy [001] axis, these measurements demonstrate the presence of local moments at high temperatures.

The occurrence of superconductivity with $T_c$ = 80 mK in the cleanest samples of \ybal{} with $\rho_0 \lesssim0.8\mu\Omega\mathrm{cm}$ raises the intriguing question of how superconductivity develops out of this local moment system. Further evidence of novel correlated electron behaviour \cite{Nakatsuji08} is signaled by low temperature non-Fermi liquid (NFL) behaviour, with the magnetic susceptibility diverging for applied magnetic field $B\parallel{}c$, accompanied by an electrical resistivity $\rho\propto{}T^{3/2}$. Application of a small magnetic field at low temperature recovers FL behaviour with enhanced specific heat coefficient $\gamma$ (generally referred to as a heavy FL), suggesting that the $f$-electron bands hybridize with the conduction electrons and thereby contribute to the Fermi volume \cite{Hewson93}. Individual local moment behaviour is found to be recovered above an energy scale of~$T^* \sim 200$~K \cite{Nakatsuji08}. This thermodynamic and transport evidence for strongly correlated electron behaviour in \ybal{} provides impetus to study the electronic structure.

%% This covers the main points of the introduction - there a few other points
%% that could be made such as the accuracy of structure calculations and
%% implications but these are probably best made in the conclusion as
%% suggestions for future work.

In this Letter we present a study of the Fermi surface (FS) of \ybal{}, observed by measuring quantum oscillations in a range of magnetic fields. High quality superconducting single crystals with typical dimension $2\times{}0.1\times{}0.01$~mm$^3$ were grown according to the method described by Macaluso $et. al.$ \cite{Macaluso07}. Where necessary, electrical contacts were made to the crystals using Dupont
6838 silver epoxy, yielding contact resistances of $\sim0.25\:\Omega$ at room
temperature.

A study of the temperature, field and angular dependence of quantum 
oscillations was carried out using a number of techniques. Low field Shubnikov-de Haas (SdH) measurements were carried out in Cambridge between 10 and 18~T at temperatures as low as 10 mK, using excitation currents 40 $\mu$A below $100$~mK to avoid sample heating. The angular dependence from $B\parallel$ [001] to [100] and $B\parallel$ [001] to [010] was studied at low fields. High field measurements were carried out using the 45~T hybrid magnet facility at the National High Magnetic Field Laboratory in Tallahassee, where oscillatory signals were observed in the field range 35 - 45~T, to temperatures as low as 40~mK. Rotations from $B\parallel$ [001] to [100] were studied using a capacitance based torque magnetometer to observe the de Haas-Van Alphen (dHvA) effect, while rotations from $B\parallel$ [001] to [010] were observed using the SdH effect. The DC magnetization was measured using a pulsed field at the ISSP in Tokyo.

\begin{figure}
	\centering
		\includegraphics[width=0.5\textwidth]{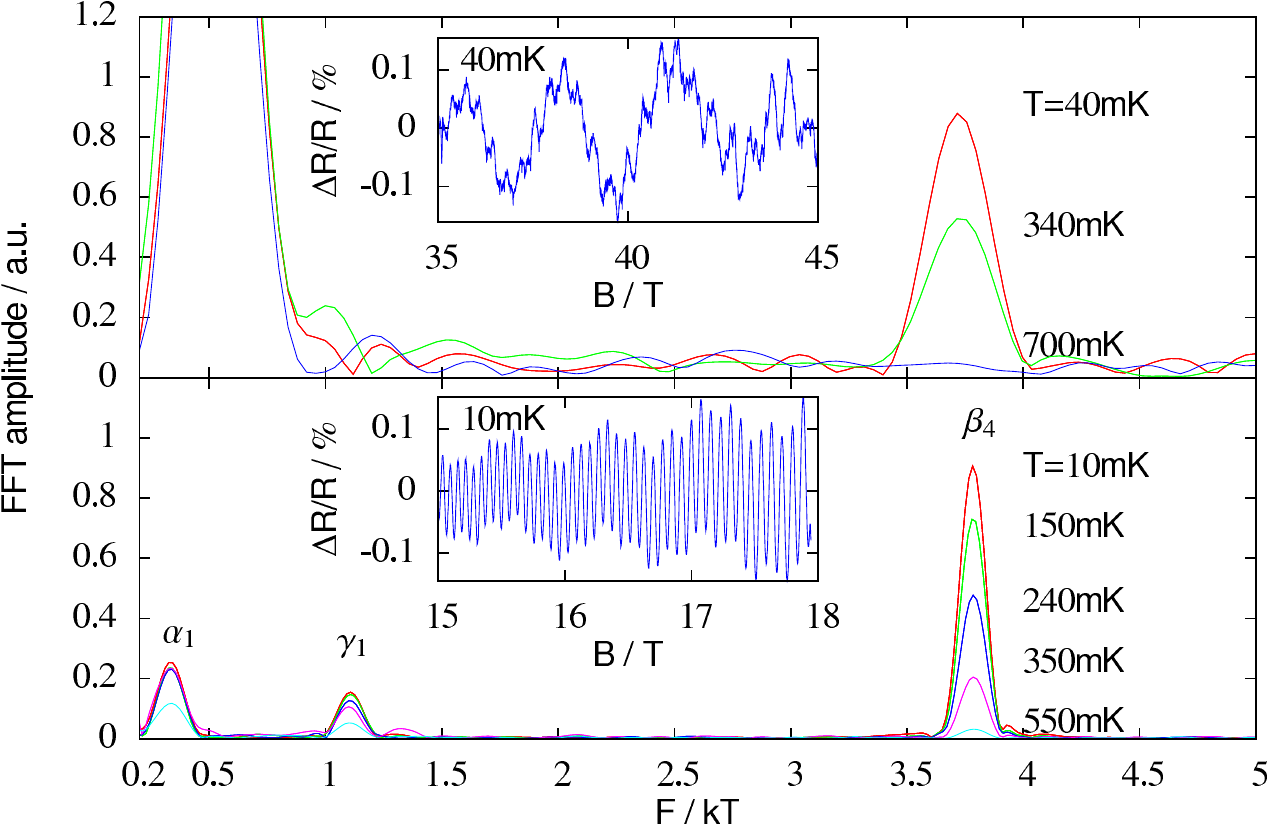}
	\caption{(color online)\emph{Inset}: Percentage change in the sample resistance as a function of field.\emph{Top panel}: SdH data obtained from 35 - 45~T B$\parallel[010]$ and associated $1/B$ FFT spectra at various temperatures. \emph{Lower panel}: SdH data obtained from 15 - 18~T B$\parallel[010]$ and associated $1/B$ FFT spectra at various temperatures.}
	\label{fig0}
\end{figure}

Fig.~\ref{fig0} shows sample oscillations in raw data treated with a polynomial background subtraction, the inset of the upper (lower) panel corresponds to high (low) field measurements with $B\parallel{}[010]$. The main panels show Fourier transforms of the data at various temperatures, with several peaks clearly visible in the spectrum associated with extremal orbits of the FS. Estimates of the quasiparticle mean free path $l_{free}$ were obtained from field dependent amplitude fits to the Dingle factor for impurity scattering $R_D=\mathrm{exp}(C_r/2l_{free})$ (where $C_r$ is the circumference of the associated real space orbit). Fitting yield $l_{free}=3400$\AA{} for orbits in the basal plane and $l_{free}=460$\AA{} for orbits perpendicular to [100], assuming circular orbits. The mean free path is of comparable length to that calculated by more indirect means \cite{Kuga08}, placing \ybal{} in the clean limit for superconductivity.

In order to clearly elucidate the role of the $f$-electrons in \ybal{} experimentally measured quantum oscillations were compared with predictions from band theory for the $4f^{14}$ case of \lbal, where the $f$-shell is filled and falls below the Fermi level, and for the $4f^{13}$ case of \ybal, where the $f$-hole is believed to be itinerant. \textit{Ab initio} electronic structure calculations were performed using the Wien2K package \cite{Blaha01}. Calculations were conducted under the local density approximation including the effect of spin-orbit coupling (LDA+SO) using full potentials and a LAPW basis. A grid of $39\times39\times64$ $\mathbf{k}$-points was used for the Brillouin zone integration. For completeness a ferromagnetic LDA+SO+U calculation with the moment along c-axis has also been performed for \ybal, yielding essentially the same FS as that found for \lbal.

\begin{figure}
	\centering
	\includegraphics[width=7cm]{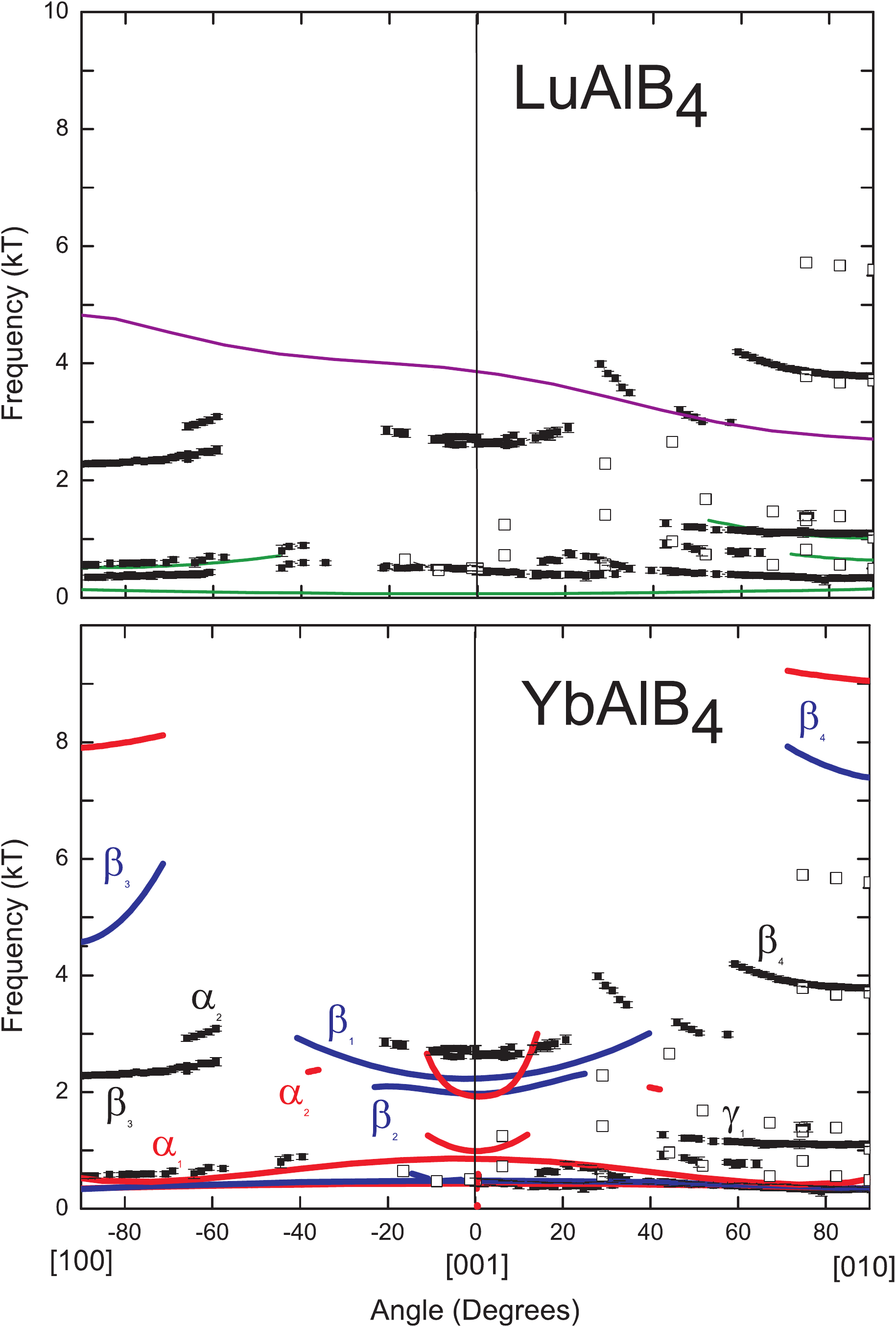}
	\caption{(color online)Angular dependence of experimental quantum oscillation frequencies at low (filled 	black square) and high (unfilled black squares) field compared to that of the LDA+SO FS:\\
	\emph{Top panel}: \lbal. Green$\to$band 91 and purple$\to$band 93
	\emph{Bottom panel}: \ybal. Red$\to$band 89 and blue$\to$band 91\\}
	\label{fig:LDASOFreq}
\end{figure}

\begin{figure}
	\centering
	\includegraphics[width=7cm]{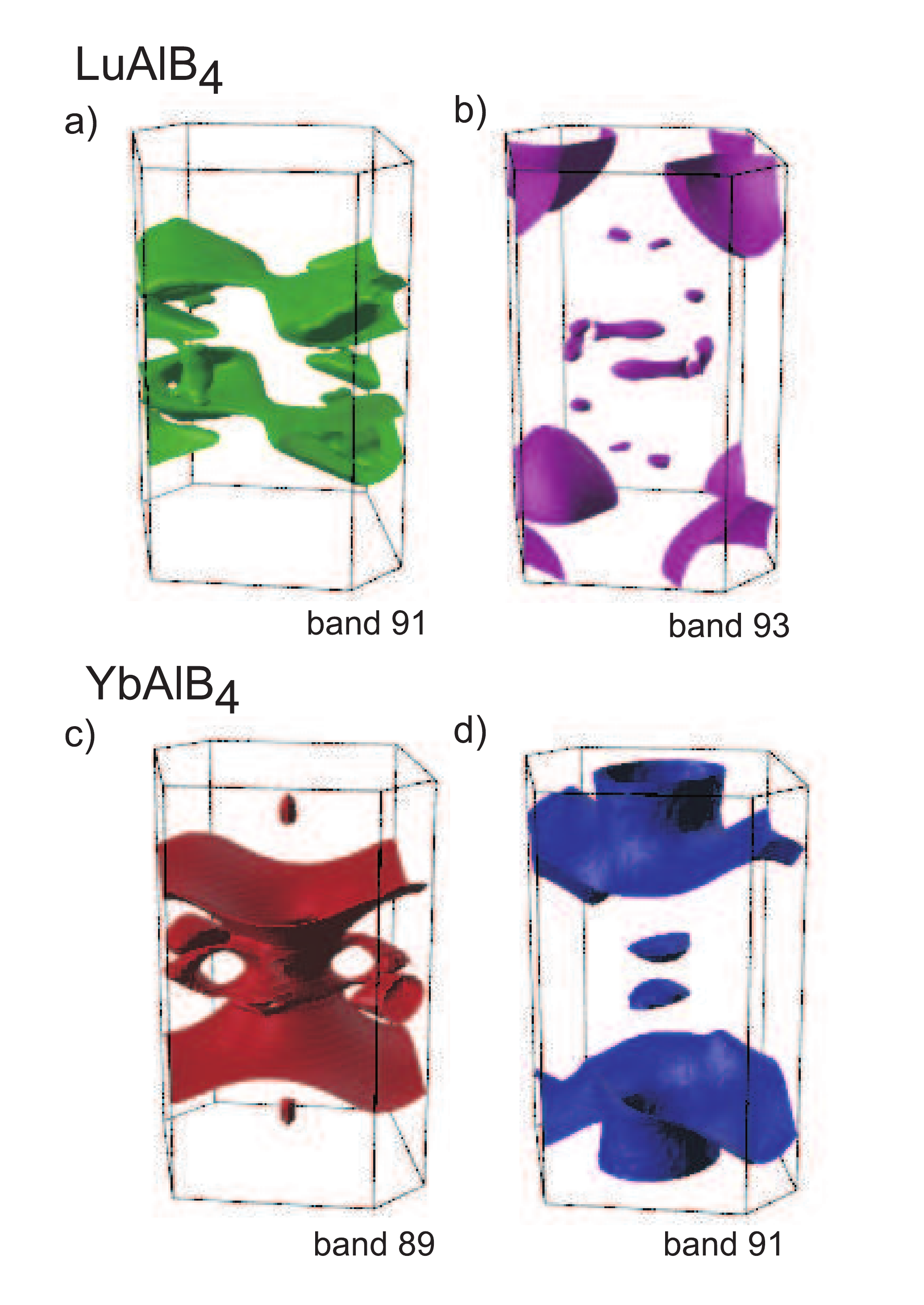}
	\caption{(color online) Panels a) and b) show the FS
          corresponding to bands 91 and 93 in \lbal{} (with a filled $f$ shell),
          calculated using the LDA technique described in the text. Panels c)
          and d) show analogous calculations for bands 89 and 91 in \ybal,
          assuming itinerant $f$-electrons. In all cases effects of
          spin-orbit coupling were included. Quantum oscillation frequencies correspond to the area enclosed by extremal orbits, for example we ascribe $\beta_1$ to the belly of band 91 around the Z point. Two additional small pockets from band 93 of \ybal in the plane of Z are not shown in the figure.}
	\label{fig:FS}
\end{figure}

Fig.~\ref{fig:LDASOFreq} shows a comparison of the angular dependence of the oscillation frequency of our experimental data at high and low fields plotted alongside the angular dependence of the LDA+SO FS for \lbal{} (top panel) and \ybal{} (bottom panel).  Fig.~\ref{fig:FS} shows the calculated FS for the main bands that cross the Fermi level for both the \lbal{} and \ybal{} cases. A strongly 3D FS is predicted by the LDA+SO calculations despite the apparent 2D nature of the crystal structure, due to strong orbital hybridization in the short c-axis direction of the unit cell.

\begin{table}
\caption{\label{tab:table1}Experimentally observed orbits and masses
  determined by fitting to
  the Lifshitz-Kosevich relation for a single mass. Together with the band mass of the assigned orbit where $\alpha$ corresponds to band 89 and $\beta$ to band 91 where an assignment was possible, $\gamma$ corresponds to experimental frequencies that are not assigned.}
\begin{ruledtabular}
\begin{tabular}{lllll}
Orbit&m$^*$ (18-15 Tesla)&Band mass&Enhancement\\
\hline
$\alpha_1$&3.6&0.25&14.4\\
$\alpha_2$&13.1&1.01&13.0\\
$\beta_1$&9.3&1.32&7.0\\
$\beta_2$&7.6&1.30&5.8\\
$\beta_3$&3.9&1.48&2.6\\
$\beta_4$&10.2&2.22&4.6\\
$\gamma_1$&5.7&-&-\\
\end{tabular}
\end{ruledtabular}
\end{table}

The lower panel of Fig.~\ref{fig:LDASOFreq} corresponds to the itinerant hole case, and shows a more satisfactory match between experiment and theory. In particular there is reasonable agreement with the experimental frequencies observed with \emph{B} $\parallel$ [001], as well as for the frequencies below 500~T observed throughout the rotation. The inclusion of the spin-orbit interaction was essential for obtaining quantitative agreement along the c-axis, which is indicative of the importance of the 4$f$ moment at the Fermi level. The large frequencies predicted for \emph{B} $\parallel$ [010] and \emph{B} $\parallel$ [100] are greater than those observed experimentally, yet possess similar angular dependence. The discrepancies are likely indicative of the presence of correlation effects beyond the LDA. The high field data, taken up to 45~T, shows broad agreement with the angular dependence of the low field data.

\begin{figure}
	\centering
		\includegraphics[width=0.5\textwidth]{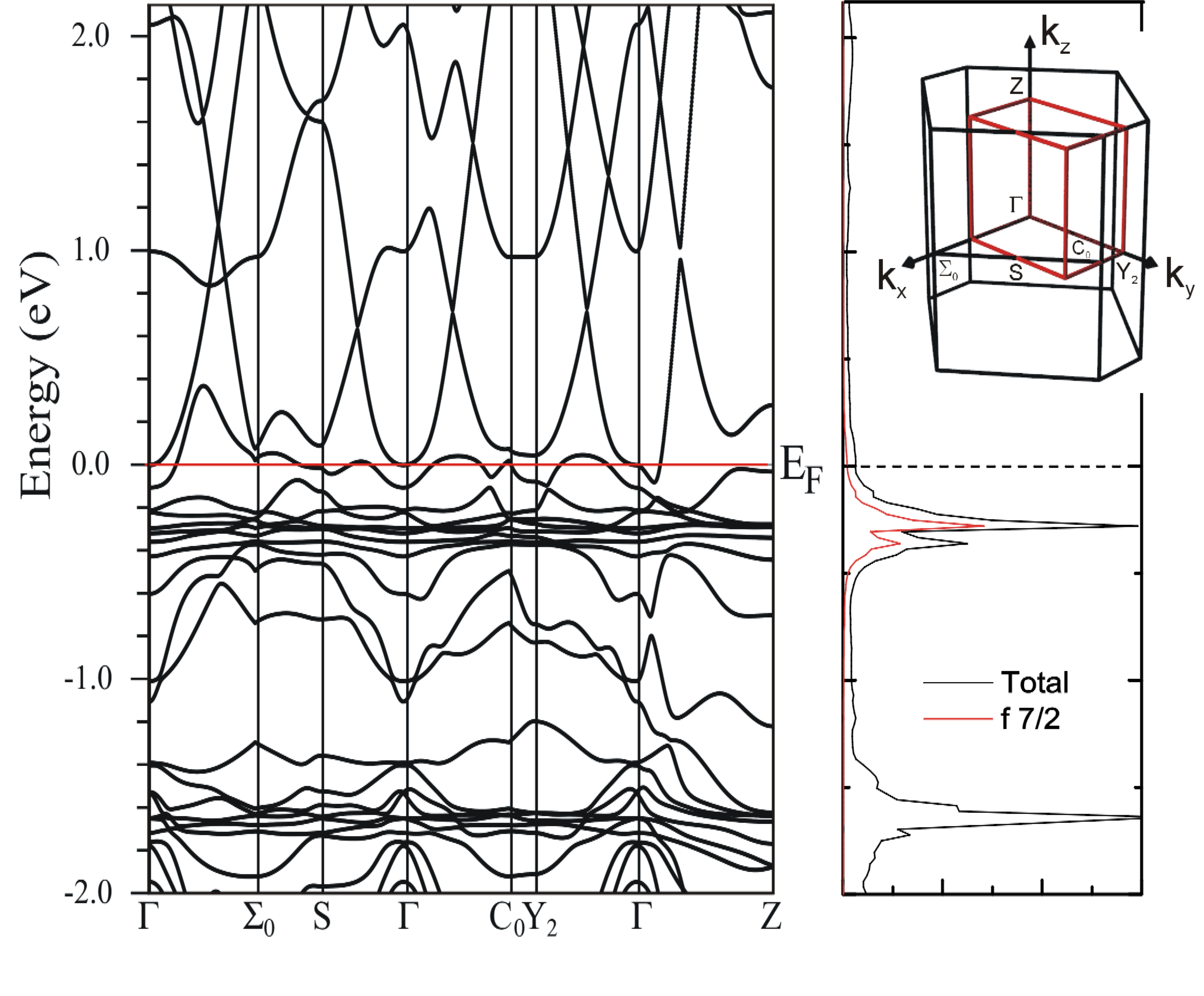}
	        \caption{Main panel (color online): Band structure produced
                  by LDA+SO. The inset
	          shows the Wigner-Seitz cell used to define the
                  crystallographic directions. Right panel:
	          The associated density of states (DOS), with the total DOS
                  in black and the contribution
	          from the $J=7/2$ states in red.}
	\label{fig:BS}
\end{figure}

The full bandstructure resulting from the LDA+SO calculation of \ybal{} along with the associated density of states is
depicted in Fig.~\ref{fig:BS}. Spin-orbit coupling splits the $f$ level into $J=7/2$ and $J=5/2$ components separated by 1.4~eV, with crystal electric field splitting further lowering the symmetry of the ground $^{2}F_{7/2}$ state, thereby yielding a ground state doublet. By summing over the projected density of states for $J=7/2$, an occupation of 7.42 was obtained, implying an $f$-hole value of 0.58. This mixed valence character yielded by the LDA+SO solution is inconsistent with the experimentally observed local moment, implying that strong coupling effects beyond the LDA are important. 

Visualization of the charge density from the itinerant calculation indicates that the predominant bonding charge density lies about the Boron layer. The projection of the Yb $J=7/2$ density of states from the full density in Fig.~\ref{fig:BS} demonstrates the strong hybridization of the Yb $f$ level with states from other atoms, in this case the Boron layer, this is of importance to the itinerant electron character of this material.

\begin{figure}
	\centering
	\includegraphics[width=0.5\textwidth]{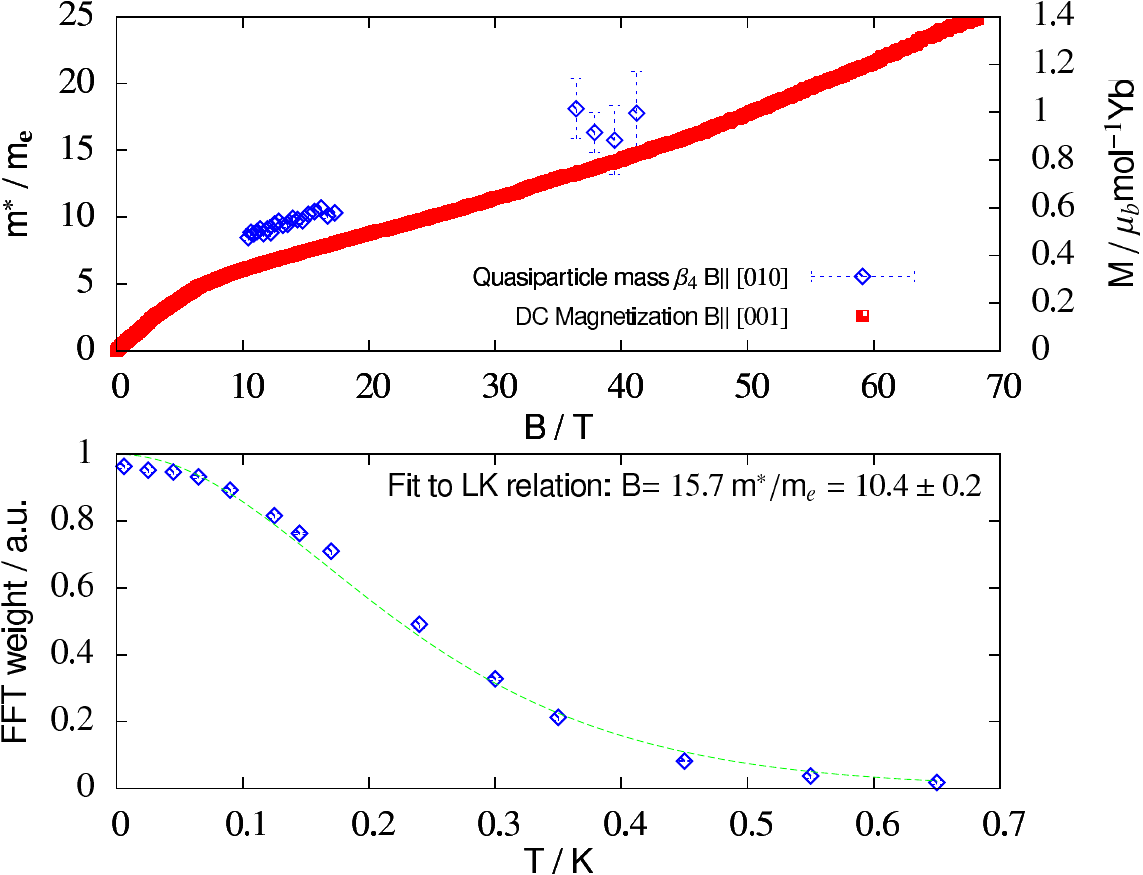}
	        \caption{\emph{Top panel}: Magnetization measured at 1.4~K $B\parallel[001]$ up to 68~T together with the field dependence of the $\beta_4$ orbit $B\parallel[010]$. Note a superlinear increase in $M$ over the range quantum oscillations were observed. \emph{Bottom panel}: Lifshitz-Kosevich fit to $\beta_4$ orbit at 15.7~T for a single mass.}
	        \label{fig:MAG}
\end{figure}

We proceed to compare the measured quasiparticle effective mass ($m^*$), extracted by fitting the temperature dependence of the FFT spectral weight to the standard Lifshitz-Kosevich relation (Fig.~\ref{fig:MAG}), with the calculated band mass. Table 1 shows a comparison of experimentally measured masses at low fields and their enhancement over calculated band masses. The sizable enhancement in effective mass is consistent with significant $f$-electron hybridization at the Fermi energy. An estimate of the contribution to the electronic specific heat component from the experimentally observed FS sheets is obtained by applying the experimental mass renormalization to the density of states calculated for the observed sheets. While the total density of states at the Fermi energy from electronic structure calculations is small, equivalent to an electronic specific heat component of $\gamma=6.4$~mJ/mol~K$^2$, taking the average of the experimental enhancement yields $\gamma\sim50$~mJ/mol~K$^2$ - comparable to the experimentally determined value of $\gamma\sim75$~mJ/mol~K$^2$ measured at 9~T. As such we are able to account for the enhancement in the specific heat, furthermore this suggests that there is no large difference in effective mass between spin channels.

The large density of states associated with hybridized bands in \ybal{} raises the possibility that the paramagnetic HFL can be polarized at modest applied magnetic fields. Polarization of the HFL would in turn lead to a shift in chemical potential, associated with a reduction in the experimentally measured effective mass. We therefore further probe the extent of $f$-electron involvement in the FS up to high magnetic fields. Fig.~\ref{fig:MAG} shows the effective mass evolution as a function of the applied magnetic field. Rather surprisingly, the modest mass enhancement persists at high magnetic fields, in fact the effective mass is seen to increase with magnetic field measured for $B\parallel[010]$. From the energy scale $E=\mu_{\mathrm{eff}}B$ associated with the highest applied magnetic fields at which quantum oscillations have been observed, we infer that $f$-electrons are involved in the Fermi surface at least up to an energy scale$\sim$100~K.

In order to understand the increase in effective mass with magnetic field, we consider the trend in magnetization at these fields (Fig.~\ref{fig:MAG}). Measurements of the magnetization up to 68~T reveal an absence of a metamagnetic transition \cite{Daou06,Aoki01}. The subtle non-linear increase in magnetization at higher fields implies a non-linear change in the density of states around the Fermi level. Accordingly, the enhancement of the effective mass shown in Fig.~\ref{fig:MAG} is consistent with an increase in the magnetic susceptibility.

Our mapping out of the electronic structure in \ybal{} could provide useful insight to the origin of unconventional superconductivity in this material. Comparing the \ybal{} system with the prototypical Ce-based unconventional superconductor CeCoIn$_5$ \cite{Petrovic01, Bianchi03}, we note potential similarities. Both these systems are located in proximity to putative quantum critical points in the vicinity of which NFL behaviour is seen; while CeCoIn$_5$ lies on the antiferromagnetic side, \ybal{} is positioned on the paramagnetic side. Our experimental measurements in \ybal{} are performed at high magnetic fields beyond the region of NFL behaviour, enabling us to access the low energy excitations of the FL prior to its evolution to the quantum critical regime. We find direct evidence that $f$-electrons participate in the Fermi surface of \ybal{} right up to temperatures of the order of 30~K, reflecting the large Kondo lattice energy scale in this material \cite{Nakatsuji08} - in fact similar to CeCoIn$_5$ \cite{Nakatsuji02}.

Given the likely association of an enhanced coherence energy scale with the appearance of superconductivity in these materials, it appears unexpected that the superconducting temperature in \ybal{} is considerably lower when compared to the Kondo temperature scale than in the case of CeCoIn$_5$. Potential clues as to the suppressed $T_{SC}$ in \ybal{} arise from the FS topology. An outstanding question relating to the appearance of unconventional superconductivity in heavy fermion systems relates to the origin of Cooper pairing from within these large moment Kondo lattice systems. Despite the large local moment in \ybal, our quantum oscillation measurements find FS pockets in which itinerant $f$-electrons are involved. We are thereby able to pinpoint the locations in reciprocal space comprising heavy electrons from which superconducting Cooper pairs likely originate. Among reasons for the vastly lower superconducting temperature in \ybal{} may perhaps be the robust 3D nature of these FS pockets: it has been suggested that the size of T$_{SC}$ is enhanced in 2D \cite{Monthoux99}.

An interesting question for further exploration involves the nature of evolution of the observed heavy FS pockets as the system is tuned toward the other side of the quantum critical point, where magnetism may potentially replace superconductivity as the groundstate. We find that an antiferromagnetic LDA+U calculation in fact yields a lower energy ground state, hinting that the system lies close to an antiferromagnetic instability. The resolution of these questions and how \ybal{} fits into the Kondo lattice phenomenology is anticipated to form the basis of future investigations.

%Thank people and grants

We thank S. K. Goh, H. Harima and G. G. Lonzarich for useful discussions, and R. G. Goodrich for the use of magnet time. This work was supported by the EPSRC, the Royal Society, I2CAM NSF grant DMR-0645461, Trinity College (Cambridge), Florida State University, JSPS (18684020) and MEXT(17071003, 19014006) of Japan. During preparation of this manuscript, we became aware of electronic band structure calculations performed by Nevidomskyy et al. \cite{Nevidomskyy08}, yielding results similar to those presented here.

\bibliography{YbAlB4_v3p5.bib}

\end{document}